\documentclass[reprint,superscriptaddress,aps,prl]{revtex4-2}
\usepackage{amsmath}
\usepackage{dcolumn}
\usepackage{graphicx}
\usepackage{amsmath,amssymb,amsfonts}
\usepackage{braket}
\usepackage{url}
\usepackage{dsfont}
\usepackage[ruled,vlined]{algorithm2e}
\usepackage{algpseudocode}
\usepackage{appendix}
\usepackage{hyperref}

\makeatletter
\newcommand*{\addFileDependency}[1]{
  \typeout{(#1)}
  \@addtofilelist{#1}
  \IfFileExists{#1}{}{\typeout{No file #1.}}
}
\makeatother

\graphicspath{figures/}

\begin{document}

\title{Stigmergic optimal transport}

\author{Vishaal Krishnan}
\affiliation{School of Engineering and Applied Sciences, Harvard University, Cambridge MA 02138.}
\author{L.~Mahadevan}\thanks{Corresponding author. lmahadev@g.harvard.edu}
\affiliation{School of Engineering and Applied Sciences, Harvard University, Cambridge MA 02138.}
\affiliation{Department of Physics, Harvard University, Cambridge MA 02138.}
\affiliation{Department of Organismic and Evolutionary Biology, Harvard University, Cambridge MA 02138.}

\begin{abstract}
Efficient navigation in  swarms often relies on the emergence of decentralized approaches that minimize traversal time or energy. Stigmergy — where agents modify a shared environment that then modifies their behavior - is a classic mechanism that can encode this strategy. We develop a theoretical framework for stigmergic transport by casting it as a stochastic optimal control problem: agents (collectively) lay and (individually) follow trails while minimizing expected traversal time. Simulations and analysis reveal two emergent behaviors—path straightening in homogeneous environments and path refraction at material interfaces—both consistent with experimental observations of insect trails.  
While reminiscent of Fermat's principle of global optimization, our results show how local, noisy agent–field interactions can give rise to geodesic trajectories in heterogeneous environments, without centralized coordination or global knowledge, but through an embodied slow–fast dynamical mechanism.
\end{abstract}

\maketitle

Efficient task execution by biological and artificial systems in complex spatial domains is central in such contexts as collective building, foraging, thermoregulation, and navigation. This typically occurs through decentralized decision-making, since the scale and complexity of the task preclude centralized control. Striking examples are seen in the architectures of social insects—from termite mounds and bee hives to ant nests—and their robotic analogs. In all these systems, coordination arises via the indirect communication of agents through environmental modification. Stigmergy (from the Greek: stigma - mark, ergos - work)~\cite{grasse1959}, reduces coordination complexity by providing a natural form of embodied memory: the actions of individuals leave persistent traces that bias subsequent actions by others, turning the environment itself into a communication channel that acts as a distributed memory coupling individual actions across space and time \cite{Bonabeau1999,Camazine2001}.  
Such mechanisms underlie trail formation, foraging, construction in termite mounds~\cite{ocko2019morphogenesis}, and  ant colonies and other biological collectives as well as their robotic analogs~\cite{prasath2022dynamics}, and exemplify a broader class of self-organized behaviors in which global structure emerges from local feedback. More generally, stigmergic coordination can be viewed as a manifestation of collective behavior arising from simple interaction rules, and admits a natural interpretation within the  physical framework of active matter~\cite{MCM-etal:13}, where energy-consuming agents interact via fields they themselves generate. Related principles also appear in distributed control and collective intelligence, spanning biological groups and engineered multi-agent systems \cite{Sumpter2006,Couzin2009}.


Classic studies have modeled trail reinforcement and nonlinear feedback~\cite{goss1989self, deneubourg1990, DJTS-MB:03, AP-BG-SG-SCN-ML-GT-VF-DJTS:12}, while others have highlighted emergent geometric regularities experimentally as being reminiscent of Fermat’s principle of least time~\cite{AMB:93, oettler2013fermat, JC-HL-WS-HC-HYK-SL-PGJ:20}.  However, a unified quantitative framework coupling agent dynamics, environmental memory, and collective optimization has remained elusive. Here we develop such a framework by casting stigmergic transport as a stochastic optimal control problem, in which agents collectively lay and individually follow trails while minimizing expected traversal time. We show that this leads to local rules for agent behavior that can achieve global optima without ever needing to know or evaluate global quantities.   

\begin{figure}[t]
\centering
\includegraphics[width=0.95\linewidth]{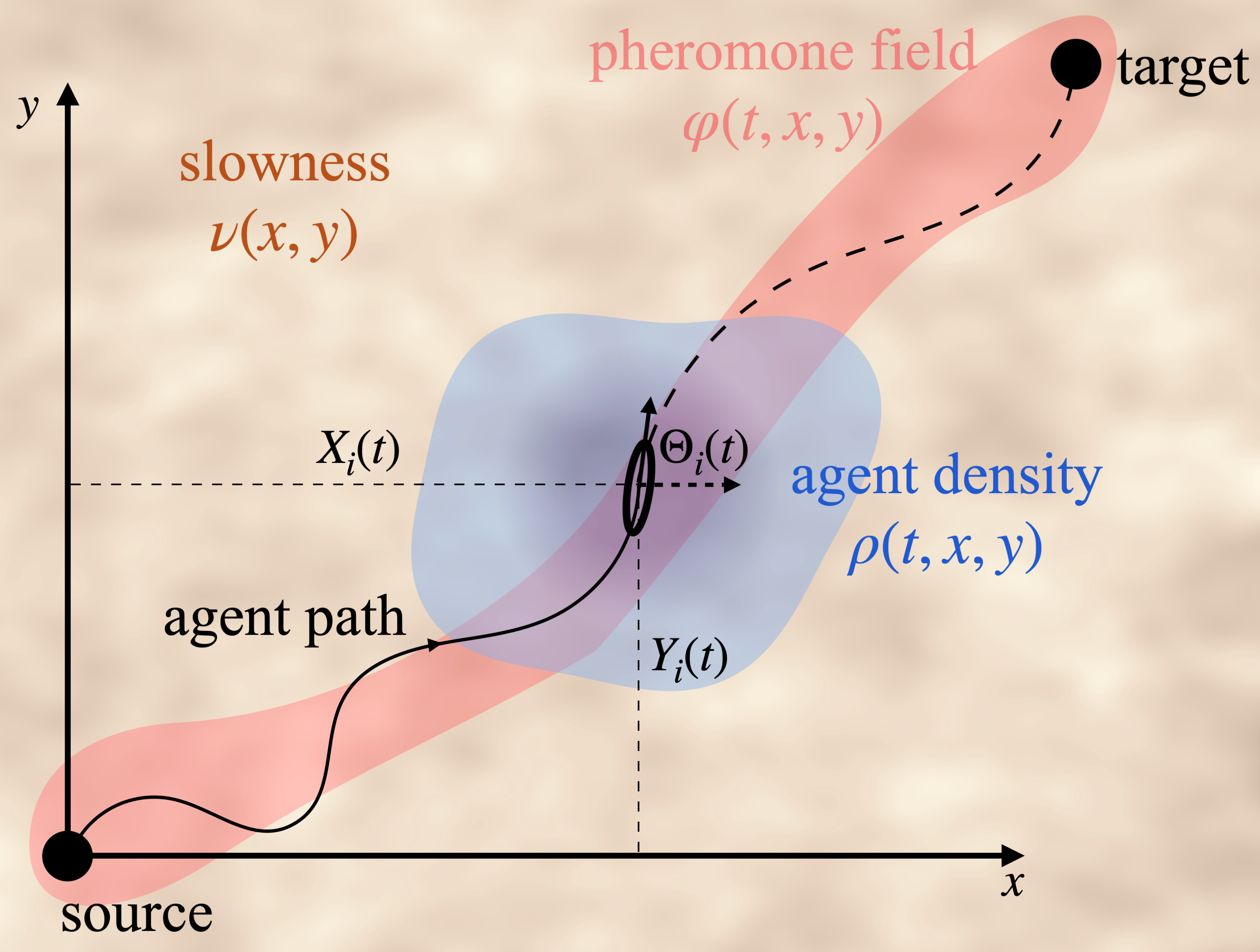}
\caption{\textbf{Stigmergic navigation and transport.} 
Schematic of collective trail following, where an individual agent~$i$, characterized by its position and orientation $(X_i(t), Y_i(t), \Theta_i(t))$, approximately follows a pheromone trail~$\phi(t,x,y)$ laid by conspecifics. At a collective level, the density field $\rho(t,x,y)$ of all agents modulates and is in turn modulated by the pheromone field $\phi(t,x,y)$.  
}
\label{fig:stigmergy}
\end{figure}

\paragraph{Trail following by agents.}
We model stigmergic transport in terms of $N$ self-propelled agents moving in a 2-dimensional domain and interacting with a spatiotemporal pheromone field $\phi(t,x,y)$. The field encodes deposited chemical signals and is locally sensed by the agents. Its gradients provide a stigmergic memory that persists across time and space, enabling coordination without direct agent–agent communication. 
Each agent $i$ is represented as an active Brownian particle with position $(X_i,Y_i)$ and orientation $\Theta_i$, obeying
\begin{align}  \label{eq:langevin}
    \begin{aligned}
        \left[ \begin{matrix} \dot{X}_i(t) \\ \dot{Y}_i(t) \end{matrix} \right] &= \frac{v_0}{\nu(X_i(t),Y_i(t))} \left[ \begin{matrix} \cos \Theta_i(t) \\  \sin \Theta_i(t) \end{matrix} \right] , \\
        \varepsilon_{\theta} \dot{\Theta}_i(t) &= \omega_i(t) + \sqrt{2 \varepsilon_{\theta} D_\theta}\,\xi_i(t),
    \end{aligned}
\end{align}
where $v_0$ is the nominal speed (which we henceforth set to 1), $\nu(x,y) > 0$ is a spatial refractive index (or equivalently, the slowness) that modulates local traversal speed, $D_\theta$ is the angular diffusion coefficient, $\xi_i(t)$ is Gaussian white noise, and $\omega_i(t)=\omega^{\mathrm{tf}}_i(t)+\omega^{\mathrm{ctrl}}_i(t)$ is the agent’s angular control input that includes both a trail-following component $\omega^{\mathrm{tf}}_i(t)$ and a trail-optimizing component $\omega^{\mathrm{ctrl}}_i(t)$ (see SM C.1 for an equivalent hydrodynamic formulation). We assume that all times are scaled by the time required to traverse the distance from the source to target, i.e. $L/\nu$ in a homogeneous system. Assuming further that the angular dynamics is relatively fast compared to this implies that $\varepsilon_\theta\ll 1$, consistent with observations~\cite{oettler2013fermat}. 

Unlike in usual theories for active Brownian particles~\cite{romanczuk2012active} where the angular velocities $\omega_i(t)$ are assumed to be known, here we need to determine a steering control law for trail-following control. We might expect that this corresponds to aligning each agent’s heading with the normalized gradient of the local pheromone field $\phi(t,X_i(t),Y_i(t))$, and in fact we can show (see SM C.2 for how this control follows via gradient descent on the K-L divergence between the agent density and pheromone concentration), that this is indeed the case, so that
\begin{equation}
\omega^{\mathrm{tf}}_i(t) = \beta \,\nabla \log \phi(t,X_i(t),Y_i(t)) \cdot 
\left[ \begin{matrix}
-\sin \Theta_i(t) \\[4pt] \cos \Theta_i(t)
\end{matrix} \right], \label{eq:trailfollow}
\end{equation}
where $\beta$ in~\eqref{eq:trailfollow} is a control gain that determines responsiveness \footnote{The dependence on~$\nabla \log \phi$ ensures scale invariance: agents respond to relative, rather than absolute, pheromone levels, which is consistent with Weber–Fechner-type sensory processing observed in biological systems.}. The law~\ref{eq:trailfollow} ensures monotonic alignment without requiring agents to estimate global geometry. 

The pheromone field $\phi(x,y,t)$ is of course dynamic, and a result of deposition by agents. We assume that it evolves according to a simple reaction-diffusion equation
\begin{align} \label{eq:pheromone_dyn}
  \partial_t \phi 
= D_\phi \nabla^2 \phi + k_{+}\,\rho - k_{-}\,\phi, 
\end{align}
where $\rho(x,y,t)$, is the Eulerian density of the agents (see SM C.3 for the dimensionless form). The two-way coupling, embodied in ~\eqref{eq:langevin}, \eqref{eq:trailfollow} and \eqref{eq:pheromone_dyn} shows how the agents both deposit and respond to pheromones. The resulting agent–field coupling is mathematically analogous to classical chemotaxis models, where agents bias their motion using a self-modified scalar field \cite{TH-KP:09}. 

It is known that the resulting reinforcement–decay dynamics can spontaneously select shortest paths and break symmetry between equivalent routes \cite{Beckers1992}. Observations suggest that the evolution of the pheromone dynamics is very slow compared to the time taken for agents to travel between the source and target. This separation of time scales, i.e. agents traverse the path relatively fast compared to the time over which the path itself changes is consistent with observations~\cite{oettler2013fermat}, and enables well-posed optimization of agent controls while ensuring consistent closure of the pheromone dynamics. Agents thus work  by communicating indirectly with each other through slowly varying environmental traces and lead to convergent algorithms for efficient steering and transport.

\paragraph{Dimensional analysis.} In the absence of control ($\omega_i=0$), agents undergo angular diffusion with persistence length $\ell_p \sim 1/(\nu_0 D_\theta)$, typically much smaller than the source–target distance $\ell_0$, making exploration inefficient. Stigmergic transport introduces a control length $\ell_{\mathrm{ctrl}} \sim \beta/D_\theta$, the distance over which angular noise is suppressed. In addition, there are two length scales associated with the environment: trail width $\sigma_{\mathrm{trail}}$, and environmental heterogeneity scale $|\nabla \log \nu|^{-1}$. The relative magnitudes of these four length scales determine whether trails are reinforced, degraded by noise, or distorted by inhomogeneities. Successful trail following requires $\ell_{\mathrm{ctrl}} \gg \ell_0$ for persistent motion, while $\ell_{\mathrm{ctrl}} \lesssim \ell_0$ leads to wandering and loss of coordination. 
 
\begin{figure}[t]
    \includegraphics[width=1.0\linewidth]{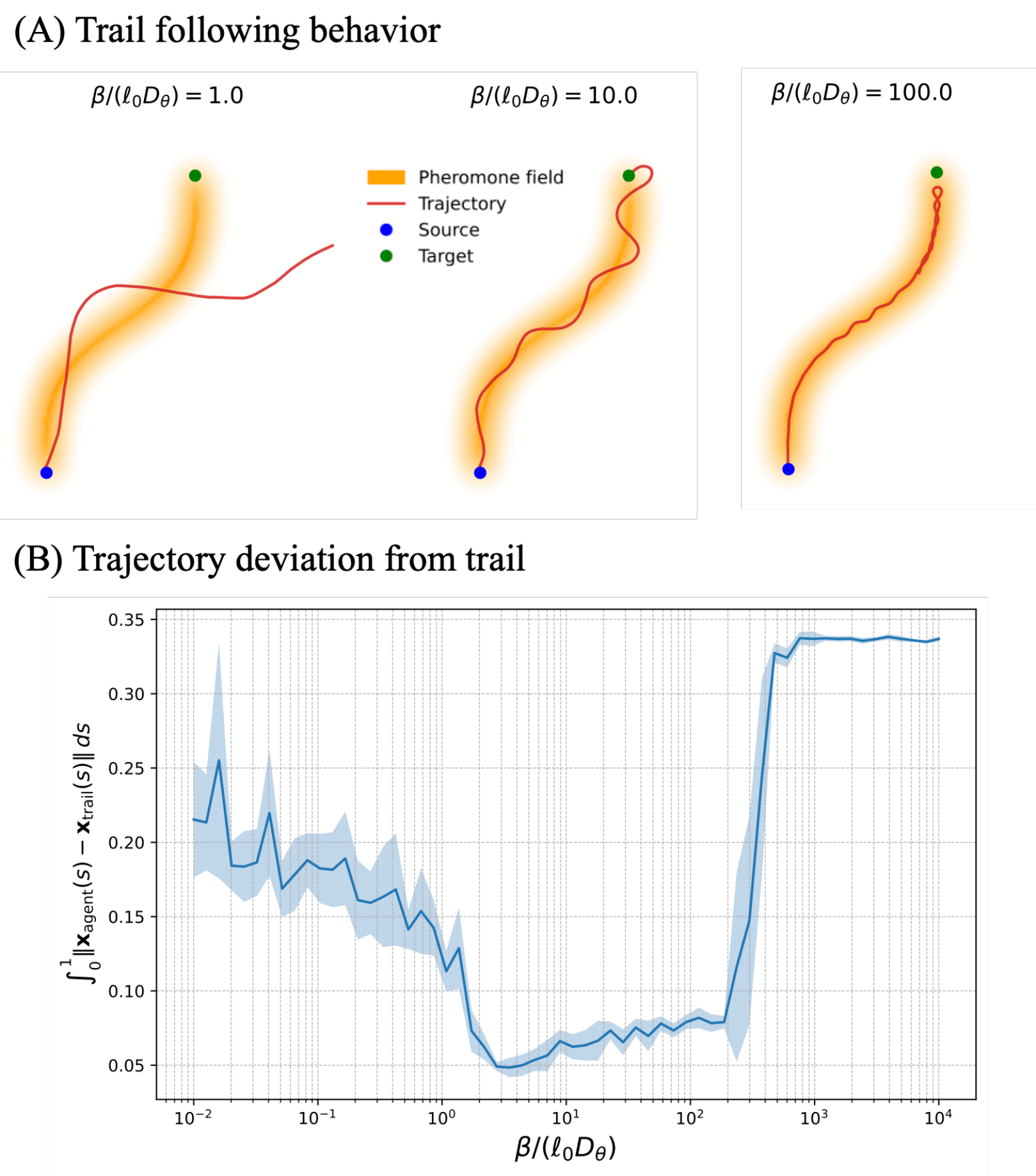}
    \caption{
    \textbf{Trail-following algorithm.} 
   (A) Representative trajectories of agents navigating from source (blue dot) to target (green dot) along a fixed pheromone trail (orange), governed by the Langevin dynamics~\eqref{eq:langevin} with trail-following control~\eqref{eq:trailfollow}, for increasing values of the dimensionless control-to-noise ratio $\beta / (\ell_0 D_\theta)$, with fixed trail sensitivity $\varepsilon = 0.1$. The shaded orange zone depicts the prescribed normalized pheromone concentration field $\phi$, ranging from white (low) to orange (high). As $\beta / (\ell_0 D_\theta)$ increases, agents transition from diffusive wandering to precise trail alignment. (B) Quantitative evaluation of trail-following accuracy as a function of the control-to-noise ratio $\beta / (\ell_0 D_\theta)$. The vertical axis reports the arc-length-averaged deviation between the agent and trail positions, $\int_0^1 \left\| \mathbf{x}_{\mathrm{agent}}(s) - \mathbf{x}_{\mathrm{trail}}(s) \right\| \, ds$, where $s \in [0, 1]$ denotes normalized arc-length. Shaded regions indicate 95\% confidence intervals over 10 stochastic trials per parameter setting. Performance improves with increasing control strength, reaching an optimal regime of accurate trail tracking. Beyond this, excessively strong control leads to over-correction and instability, degrading trail-following performance.
    }
    \label{fig:trail-following-efficiency}
\end{figure}

\paragraph{Trail optimization.}
Existing trails guide agents but require refinement, leading to a cycle: in the \emph{forward pass}, agents move from the source to target along the current trail; in the \emph{backward pass}, they return while depositing pheromone and introduce corrections that bias future trajectories. This alternation turns stigmergic reinforcement into an iterative process scheme that aims to minimize traversal time. To remove the explicit dependence on heterogeneous speeds, we introduce the arc-length parametrization $s \in [0,1]$, defined by $ds = dt/\nu(X,Y)$, exposing the geometric structure of the control problem, allowing traversal time to appear directly as a path functional (see SM C.3.2). In these coordinates, the dynamics of agent~$i$ are modified versions of ~\ref{eq:langevin} given by (see SM C.4 for reparametrization details),

{\small
\begin{align} 
    \left[ \begin{matrix} \tilde{X}'_i(s) \\ \tilde{Y}'_i(s) \end{matrix} \right] &=
    \left[ \begin{matrix} \cos \tilde{\Theta}_i(s) \\ \sin \tilde{\Theta}_i(s) \end{matrix} \right] \label{eq:langevin_reparam} \\
    \varepsilon_{\theta} \tilde{\Theta}'_i(s) &= \nu(\tilde{X}_i(s), \tilde{Y}_i(s)) \left( \tilde{\omega}_i^{\rm tf}(s) + \tilde {\omega}_i^{\rm ctrl}(s) + \sqrt{2 \varepsilon_{\theta} D_\theta} \tilde{\xi}_i(s) \right), \nonumber
\end{align}}
where $(\cdot)'=d(\cdot)/ds$. Here the unknown steering law  $\tilde{\omega}^{\rm ctrl}(s)$ is the additional component going beyond the law for trail following given by ~\ref{eq:trailfollow}, and must be determined by an extra condition, i.e. trail optimization.  This allows us to pose a stochastic control problem for the minimization of expectation of the weighted sum of the (scaled) traversal time $T = \int_0^1 \nu\big(\tilde{X}(s),\tilde{Y}(s)\big)\, ds$, and the cost associated with a corrective control $\tilde{\omega}^{\rm ctrl}(s)$, given a final state $\tilde{X}(1),\tilde{Y}(1)$, via minimizing the functional 
\begin{align} \label{eq:trail_opt}
    \begin{aligned}
    \mathcal{J} = &\mathbb{E}\!\left[
    \int_0^1 \nu\!\big(\tilde{X}(s),\tilde{Y}(s)\big)\, ds \right. \\
    & \left. + \frac{\gamma}{2}\!\int_0^1 \big(\tilde{\omega}^{\rm ctrl}(s)\big)^2 ds 
    + \Psi\big(\tilde{X}(1),\tilde{Y}(1)\big)\right],
    \end{aligned}
\end{align}
subject to the constraints imposed by the dynamics of the agents Eq. ~\ref{eq:langevin_reparam} and the evolution of the pheromone field Eq.~\ref{eq:pheromone_dyn}. Here $\gamma>0$ penalizes control effort and $\Psi$ enforces the return to the source (see SM C.3 for details). 

We solve \eqref{eq:trail_opt} using the adjoint method from optimal control theory~\cite{liberzon2011calculus}. We first introduce the adjoint sensitivities or co–states $(\mu,\Gamma)$, with $\mu(s)$ being the vector-valued Lagrange multiplier for $(\tilde{X}(s),\tilde{Y}(s))$ and $\Gamma(s)\in\mathbb{R}$ for $\tilde{\Theta}(s)$. Physically, $\mu(s)$ measures sensitivity of traversal time to local spatial deformations of the path, while $\Gamma(s)$ measures sensitivity to heading perturbations; together they encode how infinitesimal changes in geometry affect global performance (see SM C.3). Extremizing the resulting augmented Lagrangian, w.r.t. $\mu(s),\Gamma(s)$(see SM C.3 for the derivation of the co-state adjoint equations) yields the backward adjoint equations 
\begingroup\small
\begin{align} \label{eq:adjoint_backprop}
\begin{aligned}
\mu'(s) &= \nabla \nu(\tilde{X},\tilde{Y})
+ \nabla\!\left(\nu(\tilde{X},\tilde{Y})\,\left.\partial_\theta E\right|_{(\tilde{X},\tilde{Y},\tilde{\Theta})}\right) \\[-2pt]
\varepsilon_\theta \,\Gamma'(s) &= \nu(\tilde{X},\tilde{Y})\,
\left.\partial_{\theta\theta} E\right|_{(\tilde{X},\tilde{Y},\tilde{\Theta})}\,\Gamma(s)
+ \mu(s)\!\cdot\! \left[ \begin{matrix}-\sin\tilde{\Theta}\\ \;\;\cos\tilde{\Theta}\end{matrix} \right]
\end{aligned}
\end{align}
\endgroup
with terminal conditions $\mu(1)=-\nabla \Psi\!\big(\tilde{X}(1),\tilde{Y}(1)\big)$ and $\Gamma(1)=0$. Here, the potential $E(t,x,y, \theta)=-\beta \,\nabla \log \phi(t,x,y) \cdot 
\begin{pmatrix}
\cos \theta \\[4pt] \sin \theta
\end{pmatrix}$ is associated with the trail-following steering law Eq.~\ref{eq:trailfollow}. Similarly, stationarity of the augmented Lagrangian (see SM C.3) w.r.t. $\tilde \omega^{\rm ctrl}(s)$ yields the feedback law
\begin{equation}
\label{eq:ctrl}
\tilde{\omega}^{\rm ctrl}(s) = -\,\frac{1}{\gamma}\,\Gamma(s).
\end{equation}
Here $\gamma=\beta D_\theta$, a result that is analogous to the fluctuation-dissipation relation (see SM C.3) and forces the adjoint-induced flux to exactly coincide with the macroscopic momentum density. Physically this corresponds to a very intuitive rule: turn in the direction that locally decreases traversal time, with gain $1/\gamma$ (inversely proportional to the steering cost) and with spatial dependence inherited through $\nu$  and the gradient structure of $E$ (via the forward trajectory)

\begin{figure}
\includegraphics[width=1.0\linewidth]{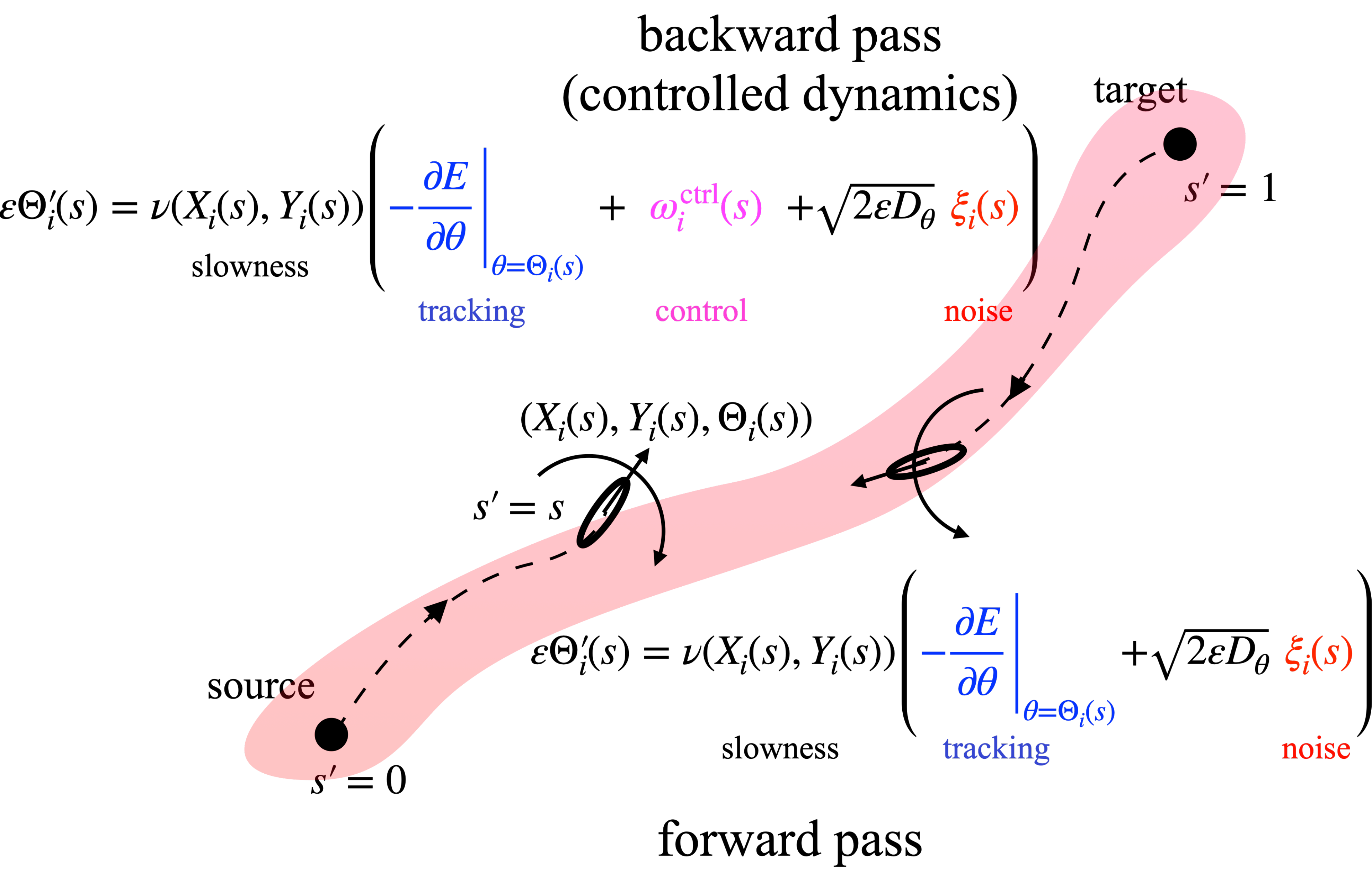}
\caption{
\textbf{Stigmergic trail laying and following.} Schematic of the path integral control procedure. To compute the trail optimizing control, we simulate \(n\) uncontrolled forward trajectories using the Langevin dynamics reparametrized by arc-length (Eq.~\eqref{eq:langevin_reparam}), each with independently sampled angular noise. The adjoint dynamics governed by Eq.~\eqref{eq:adjoint_backprop} is integrated along the forward trajectories and the trail optimizing control \(\omega_i^{\mathrm{ctrl}}\) is computed using \eqref{eq:ctrl}. For the pseudocode, see End Matter.}
\label{fig:adjoint}
\end{figure}

To determine the optimal control protocol for trail optimization, we first solve the forward problem for trail-following following $\omega^{\rm tf} = -\partial_\theta E$ given by ~\ref{eq:trailfollow} to generate a candidate path. We then solve  
for the adjoint sensitivities (co-states) by backward propagation in $s$ according to ~\ref{eq:adjoint_backprop}. Finally, we determine the trail-optimizing law Eq.~\eqref{eq:ctrl}.  Simultaneously, we  alter pheromone deposition according to Eq. (3) given the agent density. Stigmergic optimal transport is thus  a stochastic exploration of alternative paths and exploitation through pheromone reinforcement summarized as an iterative loop in Fig.~\ref{fig:adjoint}. This fast–slow loop refines trails, reduces traversal time while preserving stable trail-following, as shown in  (see SM E for details, and End Matter for the pseudocode). The process can be recast in terms of a Lyapunov functional  that guarantees convergence of the refinement loop (see SM E).

A geometric interpretation of the results from stochastic control theory suggests that trails can be interpreted as emergent geodesics of the refractive metric $ds_\nu = \nu(x,y)\, ds$. Then, curvature-driven relaxation law corresponds to motion by mean curvature, guiding convergence to time-optimal paths. From this perspective, stigmergic dynamics form a decentralized algorithm for geodesic discovery, with pheromone feedback implementing gradient descent on the traversal-time functional. An exposition of this viewpoint is provided in the End Matter.

%
 


\paragraph{Numerical simulations.} We now turn to corroborate our theory using numerical experiments to illustrate the stigmergic optimization loop shown in Fig. 4. In homogeneous environments with constant $\nu$, initially curved trails straighten over refinement cycles, driven by curvature relaxation while preserving connectivity, Fig. 4(a) (see SM C.3 for convergence results). In heterogeneous environments with piecewise-constant $\nu(x,y)$, trails refract sharply
at interfaces, converging to time-minimizing paths, Fig. 4(b), leading to Snell's law with $\nu_1 \sin \theta_1 = \nu_2 \sin \theta_2$. In both cases, we see the emergence of global geometric optimality via local stigmergic rules.
Convergence is sharpest for $\beta/(\ell_0
D_\theta)\!\approx\!1$, where exploration and exploitation are balanced. For small values of this scaled gain, i.e. weaker following capacity, agents diffuse without consolidating trails. For larger values of 
the gain, they prematurely reinforce suboptimal detours.  
\begin{figure}[!htbp]
    \centering
    \includegraphics[width=1.0\linewidth]{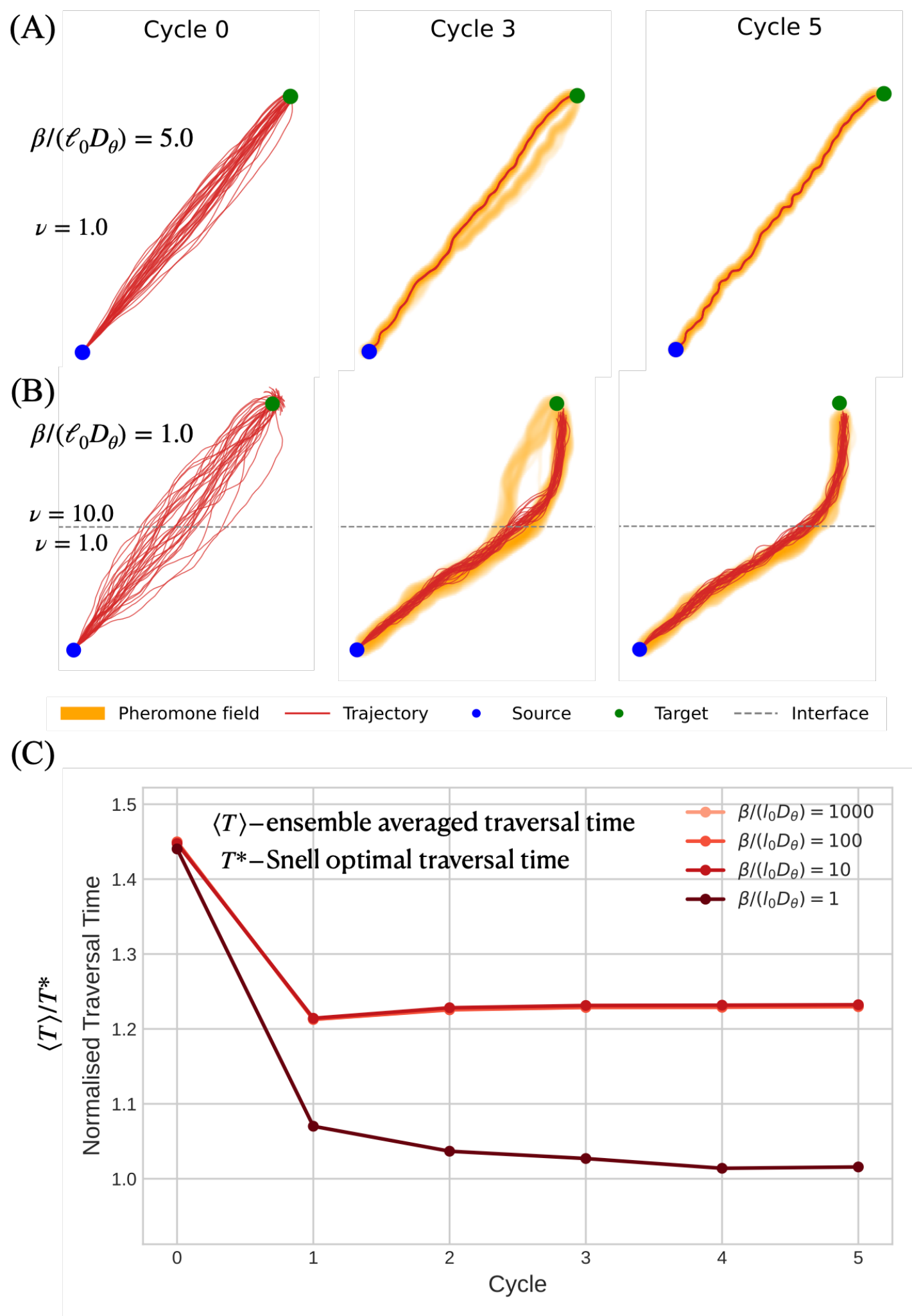}
    \caption{
    \textbf{Convergence to Snell-optimal trails and exploration–exploitation tradeoff.} Pheromone-guided refinement across stigmergic cycles. Red curves: agent trajectories; orange colormap: pheromone field. Agents move from source (blue) to target (green), depositing pheromone en route. 
    \textbf{(A)} Homogeneous medium ($\nu=1.0$): trajectories converge to the straight-line optimal path. 
    \textbf{(B)} Two-medium environment with interface (dashed line; $\nu=1.0$ below, $\nu=10.0$ above). With $\beta/(\ell_0 D_\theta)=1.0$, agents retain stochasticity to explore and converge to a refracted trail consistent with Snell’s law.  }
    \label{fig:snell}
\end{figure}


\paragraph{Discussion}
Our mathematical framework for stigmergy-based trail optimization shows how local agent dynamics coupled with pheromone-mediated feedback leads to the emergence of globally efficient routes. A variational principle for traversal time minimization produces a control law naturally decomposed into local alignment and correction components. Agents act simultaneously as trail followers and trail
improvers: they stabilize existing paths while introducing controlled
deviations that refine geometry. The relative magnitudes of the characteristic scales of persistence and gain sets the exploration--exploitation tradeoff and determines whether agents converge to globally optimal or locally trapped paths. 
More generally, the model provides an embodied approach that shows how physical trails laid down by moving agents act as both memory and guide, enabling a self-organizing solution for navigation, and serves to complement classical optimal control~\cite{liberzon2011calculus} or mean field game approaches~\cite{bensoussan2013mean} that assume centralized computation or abstract statistical coordination. Our approach opens directions for decentralized routing in swarm robotics, active matter, and programmable collectives, and suggests extensions to time-varying environments, heterogeneous populations, and learning-based adaptation. 

\paragraph{Acknowledgments} We thank the Simons Foundation and the Henri Seydoux Fund for partial financial support. All codes used for simulation results are available at: \href{https://github.com/vishaal-krishnan/stigmergic_OT}{https://github.com/vishaal-krishnan/stigmergic\_OT}

\bibliographystyle{apsrev4-2}
\bibliography{references}

\begin{appendices}
\setcounter{figure}{0}
\setcounter{equation}{0}
\renewcommand{\theequation}{A.\arabic{equation}}
 \renewcommand{\thefigure}{A.\arabic{figure}}

\section{End Matter}

\subsection{Eikonal geometry of navigation}
The problem of efficient navigation is fundamentally geometric: both natural and artificial systems seek paths through a heterogeneous medium that minimizes total traversal time between a source and target. This is similar to Fermat's principle in geometric optics~\cite{born2013principles}, the path chosen by a light ray between two given points follows the path that minimizes the functional
\[
\tau[\chi] = \int_0^1 \nu(\chi(s))\,|\dot{\chi}(s)|\,ds,
\]
where $\nu(x)$ denotes the refractive index or local slowness of the medium. It is well known that stationarity of the functional $\tau[\chi]$ over all paths yields a local curvature relation (see  SM D.1 for a derivation) $\kappa=\nabla\!\ln\nu \cdot \mathbf{n}$,  
showing that rays bend toward regions of higher $\nu$. For a homogeneous medium, this yields straight line paths, while for two homogeneous media separated by a sharp interface, this leads to Snell’s law $\nu_1\sin\theta_1=\nu_2\sin\theta_2$.  More generally, the path defines geodesics of the metric defined by the refractive index $ds_\nu=\nu(x,y)\,ds$~\cite{born2013principles}.  

An equivalent description can be given in terms of the value function $V(x,y)$ defined as the minimum remaining traversal time from  any point in the domain to the target $(x_{\rm target}, y_{\rm target})$. Then $V(x,y)$ encodes the geometry of minimal-time trajectories. Deterministic time-optimal control~\cite{SO-JS:88,liberzon2011calculus} then imply that $V(x,y)$ satisfies the eikonal equation
\[
|\nabla V(x,y)|=\nu(x,y), \quad V(x_{\rm target}, y_{\rm target})=0.
\]
whose characteristics coincide with the $\tau$-minimizing paths. This formulation connects minimal-time navigation to Hamilton–Jacobi–Bellman theory and provides a global description of optimal geometry (SM D.2).

The central result of our study is that optimal stigmergic transport collectively recover the same geometric principles without access to $V(x,y)$ itself.   Each agent only senses local pheromone concentrations and gradients and moves accordingly.   The pheromone distribution $\phi(t,x,y)$ integrates the cumulative effect of fast agent motion through deposition, diffusion and decay and acts as a slowly evolving memory.      Repeated fast agent traversals coupled to slow pheromone adaptation cause the pheromone level sets to align with the isochrones of $V(x,y)$. In effect, the system discovers the same refractive geometry that governs light propagation, but through the embodied feedback between local sensing, motion, and memory. In steady state, the emergent trail geometry coincides with $\tau$-minimizing geodesics of the refractive metric. In this sense, stigmergic trail optimization transforms the mathematics of optics into a physics of memory:  what a wavefront achieves through instantaneous propagation, a colony achieves through recurrent feedback.  The agents’ trajectories collectively carve out geodesics of least traversal time, realizing a natural bridge between geometric optics, hydrodynamic transport, and stochastic control in active matter.

\subsection{Hydrodynamic field theory}

This convergence can be thought of from a field-theoretic viewpoint as a collective computation that leads to a distributed local implementation of Hamilton–Jacobi dynamics. The colony collectively behaves as if it were following Fermat's principle, without ever solving a global variational problem. The hydrodynamic picture detailed in SM C bridges the microscopic steering and this emergent geometry. In the fast orienting limit, we find that the evolution of the agent density $\rho(t,x,y)$ and its momentum density $\mathbf{m}(t,x,y)$ obey the coupled equations
\[
\partial_t\rho+\nabla\!\cdot\mathbf{m}=0,
\quad
\mathbf{m}=\frac{1}{\nu D_\theta}\!
\int_{-\pi}^{\pi}p\,\omega(-\sin\theta,\cos\theta)\,d\theta,
\]
where the angular velocity $\omega$ determines how individual orientations contribute to the collective flux.  
Thus macroscopic transport of the colony is entirely governed by the local control of heading at the agent level. To understand how the agents, through local feedback with the pheromone field, generate the appropriate control $\omega$ that resembles geometric optics, we note that in the stochastic model, the heading of each agent evolves according to $\dot\Theta = \omega^{\mathrm{tf}}+\omega^{\mathrm{ctrl}}$,where $\omega^{\mathrm{tf}}$ aligns motion with the local pheromone gradient and $\omega^{\mathrm{ctrl}}$ provides a small adaptive correction.  

\begin{algorithm}[t] 
\label{alg:stigmergic}
\caption{\textbf{Stigmergic Transport via Adjoint Path Integral Control}}
\normalsize
\textbf{Input:} number of agents (sample paths) $i \in \{1,\ldots,n\}$ with initial states $(X_i(0),Y_i(0),\Theta_i(0))$. \\
\textbf{For each update step}:
\begin{algorithmic}[1]
    \State \textbf{Sample} $n$ independent realizations of the angular noise process $\xi_i(s)$ over $s \in [0,1]$.
    \State \textbf{Integrate forward} the agent dynamics (arc-length reparametrized Langevin system Eq.~\eqref{eq:langevin_reparam}) to generate $n$ stochastic trajectories $(X_i(s),Y_i(s),\Theta_i(s))$.
    \State \textbf{Integrate backward} the adjoint ODE system for sensitivities $(\mu(s),\Gamma(s))$ given in Eq.~\eqref{eq:adjoint_backprop} along each trajectory to obtain $\Gamma_i(s)=\partial_\theta \lambda$.
    \State \textbf{Compute and apply} the trail-optimizing control using Eq.~\eqref{eq:ctrl} and the local rule $\tilde \omega_i^{\mathrm{ctrl}}(s)=-\Gamma_i(s)/\gamma$.
    \State \textbf{Update} the pheromone field $\phi$ via deposition and decay, incorporating the optimized controls into the agent dynamics.
\end{algorithmic}
\end{algorithm}
In a hydrodynamic framework, the momentum field $\mathbf{m}$ induced by the feedback control becomes aligned with the gradient of an effective potential, and the slow pheromone field evolves so that its gradients increasingly approximate those of the value function $V$.  
Regions where agents frequently travel—corresponding to short traversal times—accumulate higher pheromone concentration, while rarely used routes decay.  
The system therefore performs a distributed gradient descent on the traversal-time functional:  
deposition and decay act as gain and dissipation, while diffusion provides regularization.  
Over repeated cycles, the coupled dynamics $(\rho,\mathbf{m},\phi)$ converge toward a stationary state, ensuring that trail reinforcement ceases once the geodesics of least time have been established. In homogeneous media, isotropic orientation leads to $\nabla\!\cdot\mathbf{m}=0$ and straight trajectories.  
In inhomogeneous environments, $\nu(x,y)$ varies spatially, and any systematic bias in $\omega$ translates into curvature of the macroscopic streamlines—precisely the analog of refraction in optics.

This picture situates trail optimization  within a hydrodynamic framework.  
The angular feedback control law (7) that acts at the level of single trajectories appears, after coarse-graining, as a constitutive law for the collective momentum field.  The slow–fast separation of timescales justifies treating the pheromone field as a quasi-static potential during each traversal, while on longer times the same field adapts to reflect the time-averaged flux.  
In the steady state, the distribution of the pheromone encodes a refractive metric through which agents move along geodesics satisfying
$\kappa=\nabla\!\ln\nu \cdot \mathbf{n}$, 
with the local slowness $\nu(x,y)$ determined self-consistently by the accumulated pheromone (see SM C.2, C.3).  
Thus, the macroscopic trail network that emerges from stigmergic feedback is mathematically equivalent to the set of optical rays in a refractive medium, but physically realized through cycles of motion, deposition, and adaptation.

\subsection{Lyapunov convergence to geodesic paths}

For frozen pheromone fields, traversal-time minimization is formulated as a stochastic optimal control problem for arc-length–parametrized trajectories $X(s)$, with cost functional
$$J=\mathbb{E}\!\left[\int_0^1 \nu(X(s))\,ds+\frac{\gamma}{2}\int_0^1 \omega_{\mathrm{ctrl}}^2(s)\,ds+\Psi(X(1))\right].$$
Stationarity of this functional yields adjoint equations for a value function
Stationarity of this control problem yields adjoint equations for a value function
$\lambda(s,x,\theta)$, whose angular derivative $\partial_\theta \lambda$ quantifies the sensitivity of traversal time to infinitesimal heading perturbations. This sensitivity directly generates the optimal local feedback law, $\omega_{\mathrm{ctrl}}=-(\nu/\gamma)\,\partial_\theta\lambda,$
so that control is implemented through local, state-dependent steering along individual trajectories (SM C.3). For the specific gain choice $\gamma=\beta D_\theta$
the flux induced by this adjoint-based control coincides exactly with the hydrodynamic momentum obtained by coarse-graining the underlying agent dynamics. This matching identifies the divergence of the macroscopic flux with the functional derivative of the traversal cost with respect to the pheromone field, $\nabla\!\cdot m=\delta J/\delta\phi
\quad \text{(SM C.5)}.$ As a result, when this flux is coupled to slow pheromone deposition, diffusion, and decay, the full agent–pheromone system can be written as a gradient flow for a composite functional. This structure implies the existence of a control–Lyapunov functional on the slow time scale, ensuring monotonic decrease of the expected traversal cost and convergence of the stigmergic refinement dynamics to a stationary configuration (SM C.4).

\subsection{Comparison with classical  optimal transport}

In the converged state, the pheromone field satisfies the Euler–Lagrange equation associated with the geometric functional
$\int \nu|\nabla\phi| ds$  so that its level sets align with the characteristics of the minimal traversal-time value function 
$V$, which satisfies the eikonal equation. The same control formulation admits a complementary interpretation in terms of classical optimal transport: traversal-time minimization under a Fokker–Planck constraint reduces to a dynamic formulation of optimal transport, a la Benamou-Brenier, with spatially varying cost, while in the deterministic limit the adjoint 
$\lambda$ plays the role of a Kantorovich potential satisfying $|\nabla\lambda|\le\nu$ (SM D.3). In this sense, stigmergic feedback realizes geodesic optimal transport through decentralized control and slow environmental adaptation, rather than explicit global optimization.

\end{appendices}

\end{document}